# Controllable spiking patterns in long-wavelength VCSELs for neuromorphic photonics systems

Antonio Hurtado[1,*] and Julien Javaloyes[2]

[1]Institute of Photonics, SUPA Department of Physics, University of Strathclyde, TIC Centre, 99 George Street, Glasgow, G1 1RD (United Kingdom)

[2]Departament de Fisica, Universitat de les Illes Balears, c/Valldemossa km 7.5, 07122, Mallorca, Spain

antonio.hurtado@strath.ac.uk

Multiple controllable spiking patterns are obtained in a 1310 nm Vertical Cavity Surface Emitting Laser (VCSEL) in response to induced perturbations and for two different cases of polarized optical injection, namely parallel and orthogonal. Achievement of reproducible spiking responses in VCSELs operating at the telecom wavelengths offers great promise for future uses of these devices in ultrafast neuromorphic photonic systems for non-traditional computing applications.

The emulation of neuronal responses for novel paradigms in computing forms an area of important research. Electronic implementation of neuron models has been studied for decades [1]. Photonic approaches have only recently emerged as these offer promise for ultrafast speeds, much faster than the millisecond time-scales of neurons [2-22]. One of these approaches considers the use of Semiconductor Lasers (SLs), as these devices can exhibit a variety of responses similar to those observed in neurons (e.g. excitability [23-26] and nonlinear dynamics [27-28]) but up to 9 orders of magnitude faster. SLs are also discrete components permitting their integration in high density circuits making them ideal for future optical interconnects and processing modules [29]. Hence, combining neuronal concepts with photonics technologies where SLs are at the core opens new routes for ultrafast neuromorphic photonic computing systems.

Recent works have reported theoretically on [18-22] the use of SLs for neuro-inspired photonic systems. Experimental studies have also emerged [8-17]. Neuro-inspired parallel information applications [8] and pattern formation [9-10] have been reported with a SL subject to optical feedback. Nahmias et al [7] have reported a photonic spike processing unit using a fiber laser with a saturable absorber. Generation of excitable spikes has also been reported in various laser structures [11-14] and their use for neuro-inspired photonic components was suggested. Amongst SLs, Vertical Cavity Surface Emitting Lasers (VCSELs) offer important advantages compared to in-plane devices, i.e. low costs, ease to integrate in 2D arrays, high coupling efficiency to optical fibres [30-31]. However, in spite of these features, it is only recently that VCSELs have started to attract attention for neuro-inspired photonics. Emulation of neuronal responses has been reported based on the polarization switching and dynamics induced in these devices under optical injection [15-16]. Also, firing of self-generated [32] and controllable [12] spikes has been observed in VCSELs and the use of these features for all-optical data storage has been proposed [17]. Furthermore, controllable spike firing in a VCSEL under different cases of polarized injection has been predicted [18]. Here, we report experimentally and numerically on the achievement of different controllable spiking patterns, namely single and multiple spikes and bursts of spikes, with sub-nanosecond (sub-ns) speed resolution in a VCSEL subject to either parallel or orthogonally polarized optical injection. Moreover, our approach uses inexpensive devices operating at the important telecom wavelength of 1310 nm. These results offer great potential for novel uses of VCSELs as fast and reconfigurable neuromorphic computational elements for non-traditional information processing paradigms.

Fig. 1(a) shows the setup used to inject polarized light from a tuneable laser (Master Laser, ML) into a 1310 nm VCSEL. The VCSEL had a measured threshold current ($I_{th}$) of 0.63 mA and its free-running optical spectrum (plotted in Fig. 1(b)) showed emission ~1332 nm when biased with 3mA. The VCSEL's temperature was kept constant at 293 K at all times. The two peaks in the spectrum correspond to the two orthogonal polarizations of the VCSEL's fundamental transverse mode. Throughout this work we refer to parallel (orthogonal) polarization to that of the VCSEL's main lasing peak (subsidiary mode). We investigate the injection of time-varying signals which are generated by modulating externally the ML's output with a Mach-Zehnder (MZ) Modulator and a Signal Generator. Fig. 1(c) plots a typical injected signal characterized by a constant level ($k_{inj}$) and perturbations (in the form of power drops) with controlled strength ($k_p$), temporal duration ($t_d$) and repetition rate ($f_{rep}$). $k_p$ is defined as the ratio between the power drop and the total injection strength ($k_p = k/k_{inj}$). A first polarization controller is included to maximize the power at the MZ modulator's output whilst a second one is used to set the polarization of the injected signal with either parallel or orthogonal polarization. Finally, the VCSEL's reflective output is analyzed with a 12 GHz amplified photodetector and a 13 GHz real time oscilloscope and with an Optical Spectrum Analyzer.

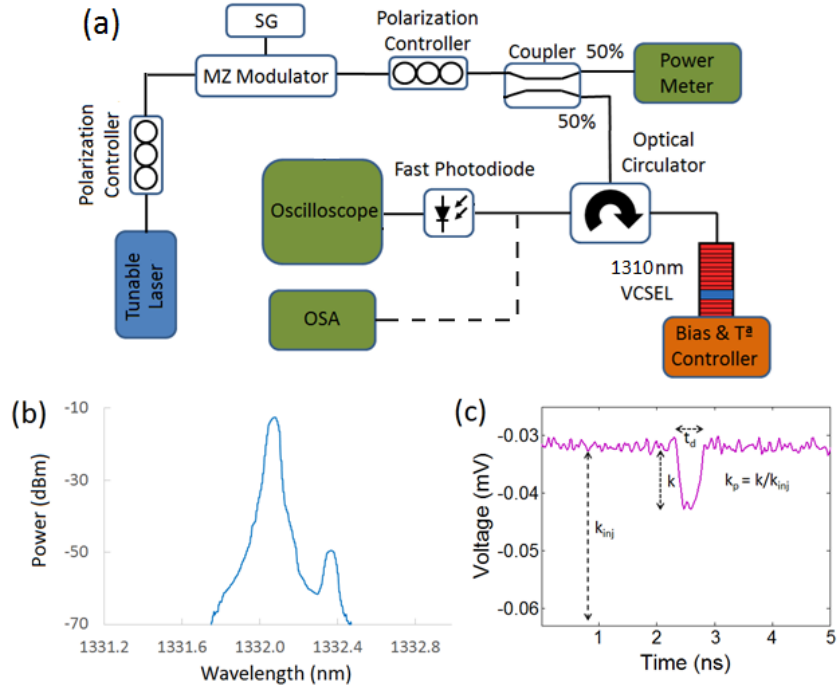

*Fig. 1 (a) Experimental Setup. (b) Spectrum of the solitary VCSEL. (c) Injected signal showing a perturbation (power drop). MZ=Mach-Zehnder, SG=Signal Generator, OSA=Optical Spectrum Analyser.*

Figs. 2(a-d) show time series (left) and temporal maps (right) measured at the VCSEL's output with the device subject to parallel (figs. 2(a) and 2(b)) and to orthogonally-polarized (figs. 2(c) and 2(d)) injection. Fig. 2's caption collects the values of the system parameters used in each case, namely bias current, $I_{Bias}$, injection strength, $K_{inj}$, frequency detuning, $\Delta f_{par/ort}$ (difference between the frequencies of the injected signal, $f_{inj}$, and the VCSEL's parallel, $f_{par}$, or orthogonal mode, $f_{ort}$: $\Delta f_{par/ort} = f_{inj} - f_{par/ort}$), perturbation's strength, $k_p$, and temporal duration, $t_d$. The level of $K_{inj}$ was enough to induce stable injection locking in both cases of polarized injection. In this situation, the arrival of a perturbation brings the system out of the locking range triggering the firing of different spike patterns depending on the initial conditions. These spikes are very similar to isolated spikes corresponding to the so-called excitable response found close to the boundary of the locking-unlocking transition [18][24]. Specifically, the time series in figs. 2(a-d) show respectively the achievement of single spiking (figs. 2(a) and 2(c)) and bursting (bursts of multiple spikes) responses (figs. 2(b) and 2(d)) following the arrival of a perturbation and when the VCSEL is subject to parallel or orthogonally-polarized injection [18]. The plots at the right side of figs. 2(a-

d) show measured temporal maps [17][33] using the repetition rate ($f_{rep}$ = *15 MHz*) as folding parameter. These maps plot superimposed time series obtained for 100 consecutive perturbations. The colour code in the maps indicates an increasing intensity from blue to red; light blue/green correspond to the steady state, dark blue indicates drops in power below the steady state and the spikes are represented in red/yellow. Figs. 2(a-d) show that the same spike patterns are obtained upon the arrival of every perturbation illustrating the controllability and repeatability of the spiking responses. These results also show the potentials of these scheme to convert rectangular signals (such as digital data signals) into trains of spikes. This feature opens the door for the use of VCSELs in binary-to-neuromorphic signal encoding elements for neuro-inspired photonic processing modules. Moreover, this system also benefits from fast operation speeds (sub-ns), low input power requirements (~tens of µW) and the use of devices operating at telecom wavelengths (1310nm) thus making this approach totally compatible with optical networks.

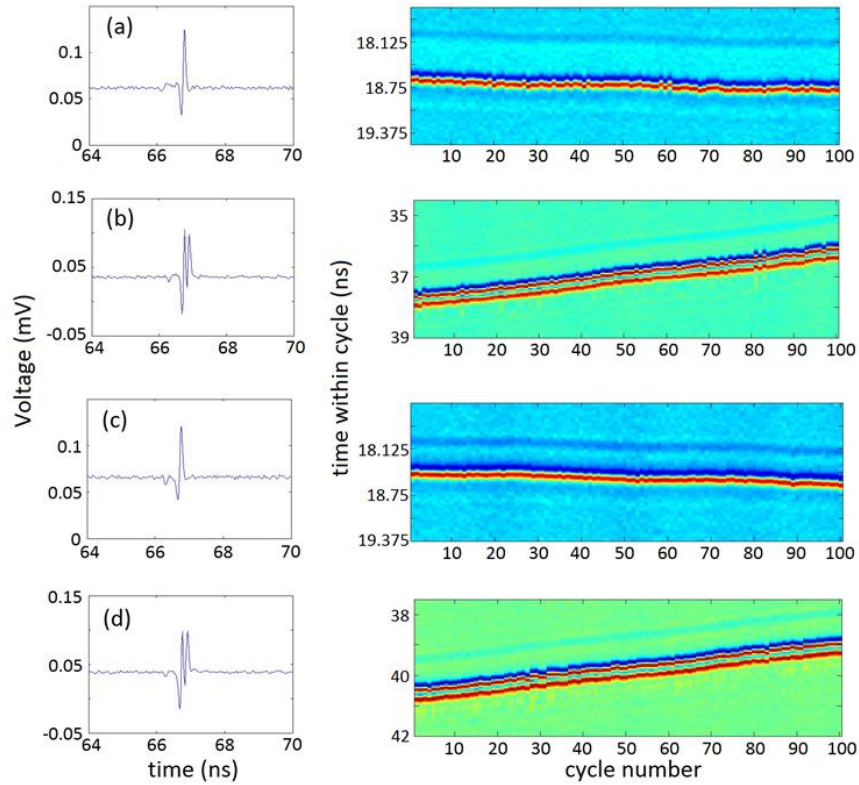

*Fig. 2. Time series (left) and temporal maps (right) of the VCSEL under parallel (a & b) and orthogonally (c & d) polarized injection showing single spiking (a & c) and bursting (b & d). ($K_{inj}$, $\Delta f_{par/ort}$, $I_{Bias}$, $k_p$, $t_d$) were set equal to: (a) (115µW, -4'65GHz, 4mA, 0.25, 0.65ns); (b) (60µW, -3.5GHz, 3mA, 0.25, 0.5ns); (c) (45µW, -3GHz, 4mA, 0.25, 0.5ns) and (d) (75µW, -3.65GHz, 3mA, 0.25, 0.5ns).*

We have also investigated the effect of the perturbation's characteristics, i.e. strength ($k_p$) and temporal duration ($t_d$), in the attained spiking responses. Fig. 3 plots measured temporal maps when $t_d$ is increased from 0.5 ns to 1.95 ns (fig. 3(a)) and to 2.65 ns (fig. 3(b)) whilst keeping constant the value of $k_p = 0.25$. Values for additional parameters, e.g. $I_{Bias}$, $K_{inj}$ and $\Delta f_{par/ort}$ are given in fig. 3's caption. Figs. 3(a) and 3(b) plot steps of 20 superimposed time traces for every case of $t_d$ studied. Specifically, figs. 3(a) and 3(b) show maps when the VCSEL is subject to orthogonally- and parallel-polarized injection. Fig. 3(a)/(fig. 3(b)) shows that at first for $t_d = 0.5$ ns a single spike/(a burst of two spikes) is obtained upon the perturbation's arrival. As $t_d$ is increased a higher number of events are produced: two for $t_d = 0.85$ ns, three when $t_d = 1.45$ ns and so on. Hence, a transition from single to multiple spiking events (single to continuous spiking/bursting) is obtained with growing $t_d$. Such response is analogous to the tonic spiking and tonic bursting dynamics in neurons [34-35] characterized by the firing of consecutive spikes (or bursts of spikes) for the whole duration of a stimulus.

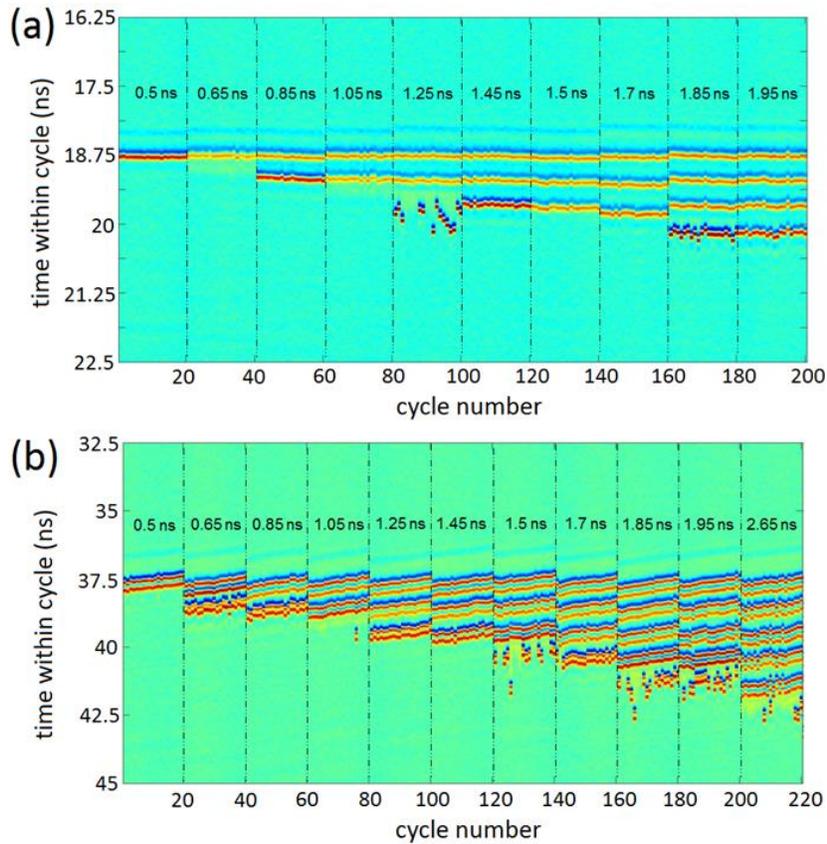

*Fig. 3. Temporal maps measured for different values of $t_d$ when the VCSEL subject to orthogonal (a) and parallel polarized (b) injection. ($I_{Bias}$, $K_{inj}$ $k_p$, $\Delta f_{par/ort}$) were set equal to: (a) (4mA, 45µW, 0.25, -3GHz); (b) (3mA, 115µW, 0.25, -4.65GHz).*

The effect of the perturbation's strength, $k_p$ was also investigated. Fig. 4 plots measured time traces and temporal maps for two different values of $k_p$ while keeping constant the rest of parameters (see caption in fig. 4 for values). The VCSEL was subject to either parallel (fig. 4(a-b)) or orthogonally (fig. 4(c-d)) polarized injection. In both cases, for $k_p = 0.15$ no significant response is obtained (see figs. 4(a) and 4(c)). However, when a perturbation of sufficient strength ($k_p = 0.25$), able to bring the system out of the locking state enters the VCSEL, a single spike is fired after every perturbation (see figs. 4(b) and 4(d)).

This is further illustrated in figs. 5(a) and 5(b) plotting temporal maps for increasing values of $k_p$ (with constant $t_d$) and with the device subject to orthogonally- and parallel-polarized injection respectively. Figs. 5(a) and 5(b) illustrate cases where different spiking regimes are obtained after a perturbation's arrival, namely single spiking and bursting. For low enough values of $k_p$ no significant changes are observed at the VCSEL's output. It is only after $k_p$ exceeds a threshold level (equal to $k_p = 0.19$ and $k_p = 0.3$ respectively for the cases of figs. 5(a) and 5(b)) a perturbation triggers a spiking response. Also, a reduced spike firing delay was measured as $k_p$ was increased [18]. Both the threshold for spiking and the different spiking delays for increasing stimuli strength are computational features also observed in neurons. These also respond firing spikes upon the arrival of stimuli of sufficient strength (remaining quiescent otherwise) [34-35] using the latency in spike firing to encode the stimulus strength [34].

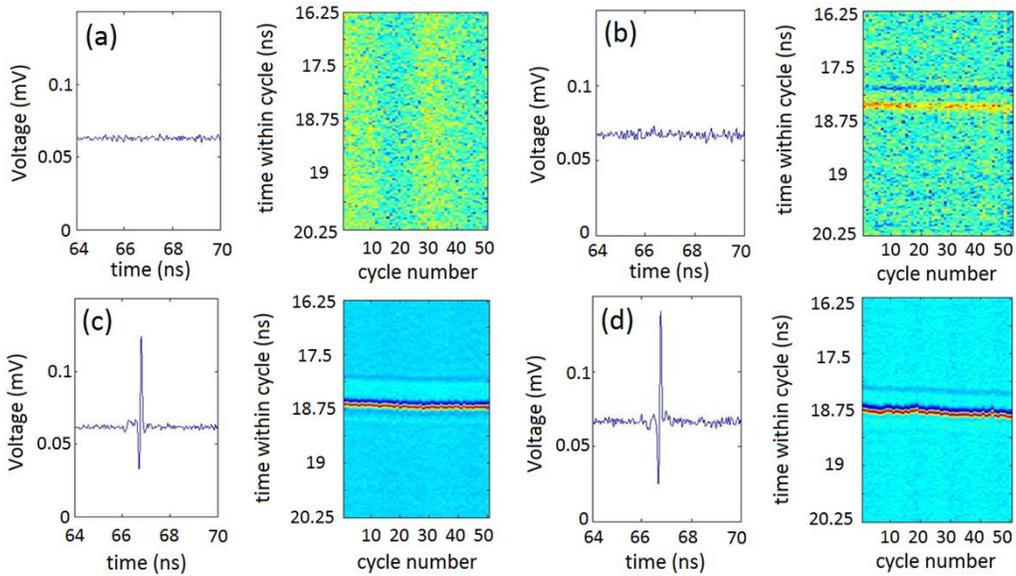

*Fig. 4. Time traces and temporal maps measured with the VCSEL subject to parallel (a & c) and orthogonal (b & d) polarized injection for two different values of $k_p$: (a & b) $k_p=0.15$; (c & d) $k_p=0.25$. ($I_{Bias}$, $K_{inj}$ $t_d$, $\Delta f_{par/ort}$) were equal to: (a & c) (4mA, 60µW, 0.65ns, -3.5GHz); (b & d) (4mA, 45µW, 0.5ns, -3GHz).*

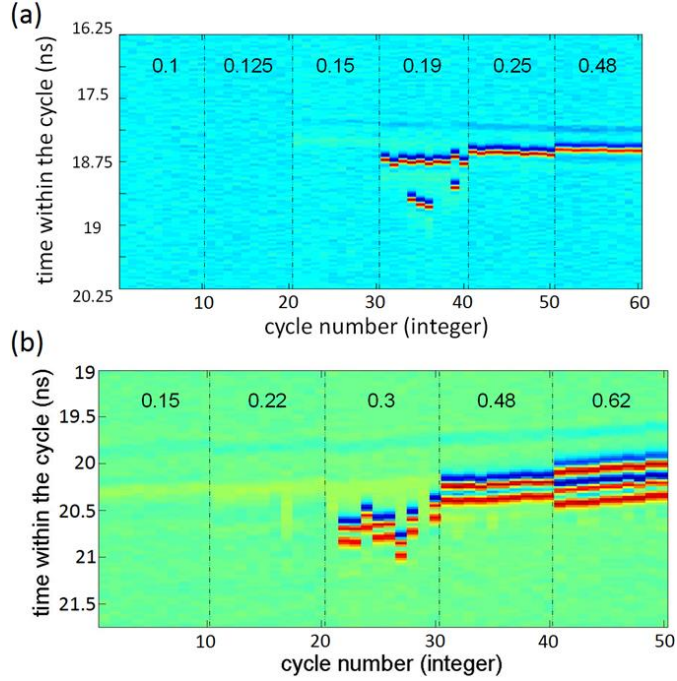

*Fig. 5. Temporal maps of the VCSEL under orthogonal (a) and parallel polarized (b) injection for different perturbation's temporal duration. ($I_{Bias}$, $K_{inj}$ $t_d$, $\Delta f_{ort/par}$) were equal to: (a) (4mA, 45µW, 0.5ns, -3GHz); (b) (3mA, 75µW, 0.5ns, -3.65GHz).*

We have also developed a numerical model to analyse the experimental results. The equation governing the evolution of the laser phase, relative to the optical injection reads as follows:

$$\dot{\Phi} = -\frac{dU}{d\Phi}, \quad U(\Phi) = -\Delta\Phi - Y(t)\sqrt{1+\alpha^2}\cos(\Phi + u)$$

where $\Delta$ is the detuning between the laser emission's frequency and the optical injection of amplitude Y; $\alpha$ is the linewidth enhancement factor and $u = \arctan \alpha$. For a steady value of the injection's amplitude, the system possesses a single stable (and an unstable) equilibrium point, provided that $Y \geq Y_c$, with $Y_c = |\Delta|/\sqrt{1+\alpha^2}$. In this case, the potential $U(\Phi)$ exhibits a minima and a maxima corresponding to these two fixed points, respectively. Temporary drops of duration T in the injection's amplitude below $Y_c$ results in a momentary unlocking of the phase and therefore of the VCSEL's emission frequency with respect to the injection field. During that interval of time, the relative phase $\Phi$ evolves in a slanted 'washboard' potential where $U(\Phi)$ has no minima. Here, the system performs periodic falls of $2\pi$ (see fig. 6(a)). We denote the duration of these $2\pi$ slips as $\tau_s$. When the duration of the power drop in $Y(t)$ is close to $T = n\tau_s$, the system finds again a stable fixed point after the *n-th* fall and will remain there. A spike is fired for every obtained 2π phase slip therefore creating a temporal trace with *n* number of spikes (see fig. 6(b)). Such a behaviour is exemplified in the cases of n = 1 and n = 3 shown in figs. 6(a) and 6(b). This is the exact same response obtained in the experiments as seen in fig. 3. However, if the amplitude of the power drop is too small (below a certain threshold) the system is not able to perform a 2π phase slip before finding again the stable fixed point and no spikes are fired. Such behaviour shown in the leftmost region of figs. 6(a) and 6(b) is identical to the experimental results included in figs. 4 and 5. Also, a similar behaviour is found if the temporal duration of the power drop is too short as compared to $\tau_s$.

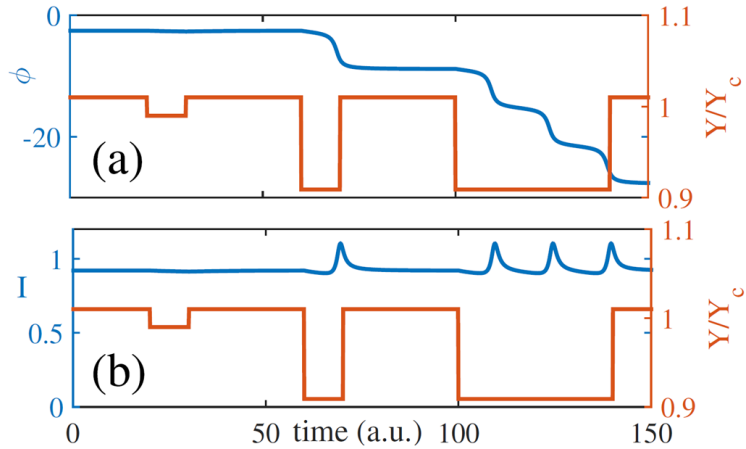

*Fig. 6. (blue) Temporal traces for the phase Φ (a) and the laser output intensity I (b). Amplitude of the injection Y/$Y_c$ (red). Numerical parameter values were set equal to: Δ=-1, α=2, $Y_c$=0.447.*

In summary, we reported on the achievement of controllable spiking patterns in a VCSEL subject to parallel and orthogonally polarized optical injection. Single and multiple spiking and bursting responses with sub-ns speed resolution are experimentally obtained in a VCSEL upon the arrival of perturbations (in the form of rectangular signals). A numerical model was also developed showing good agreement with the experiments. The achieved spiking responses exhibit strong similarities with those observed in neurons but on a much faster time-scale (over 7 orders of magnitude faster). These results added to the particular advantages of VCSELs and the use of commercial devices operating at the telecom wavelength of 1310 nm offer promise for novel neuromorphic photonic systems such as photonic digital-to-spiking information converters and photonic spiking information processing modules for use in non-traditional computing and optical networks.

**Acknowledgements**

This research received funding from the University of Strathclyde (Chancellor's Fellowships Programme). AH wishes to thank Prof. M. Adams (Essex) for useful discussions and Profs. T. Ackemann and A. Kemp (Strathclyde) for lending experimental equipment.